\documentclass[journal]{IEEEtran}
\usepackage{hyperref}
\usepackage{multirow}
\usepackage{rotating}
\usepackage{array}
\usepackage{multicol}
\usepackage{float}
\usepackage{wrapfig}
\usepackage{xr}
\usepackage{algorithm}
\usepackage{algorithmic}
\usepackage{longtable}
\usepackage{amsmath, amssymb, mathtools, amsthm, url,amsbsy,enumerate}
\usepackage{color}
\usepackage{cleveref}

\newtheorem{problem}{Problem}[section]
\newtheorem{definition}{Definition}[section]

\ifCLASSINFOpdf
\else
\fi

\begin{document}
%
\title{Simultaneous detection of multiple change points and community structures in time series of networks}
%
%
%

\author{Rex~C.~Y.~Cheung,
        Alexander~Aue,
        Seungyong~Hwang,
        and~Thomas~C.~M.~Lee,~\IEEEmembership{Senior Member,~IEEE}
\thanks{Rex~C.~Y.~Cheung is with the Department of Decision Sciences,
San Francisco State University, San Francisco, CA 94132 USA e-mail: rexcheung@sfsu.edu.}
\thanks{Alexander~Aue, Seungyong~Hwang and Thomas~C.~M.~Lee are with the Department of Statistics, University of California, Davis, CA 95616 USA e-mail: \{aaue, syhwang, tcmlee\}@ucdavis.edu}
\thanks{The authors are most grateful to the reviewers and the associate editor for their constructive and helpful comments.}
}

%
%

\markboth{}%
{Shell \MakeLowercase{\textit{et al.}}: Bare Demo of IEEEtran.cls for IEEE Journals}
%



\maketitle

\begin{abstract}
  In many complex systems, networks and graphs arise in a natural manner. Often, time evolving behavior can be easily found and modeled using time-series methodology.  Amongst others, two common research problems in network analysis are community detection and change-point detection.
Community detection aims at finding specific sub-structures within the networks, and change-point detection tries to find the time points at which sub-structures change. We propose a novel methodology to detect both community structures and change points simultaneously based on a model selection framework in which the Minimum Description Length Principle (MDL) is utilized as minimizing objective criterion. The promising practical performance of the proposed method is illustrated via a series of numerical experiments and real data analysis.
\end{abstract}

\begin{IEEEkeywords}
minimum description length, network segmentation, stochastic block models
\end{IEEEkeywords}

%
\IEEEpeerreviewmaketitle

\section{Introduction}
%
%
%
%
\IEEEPARstart{N}{etworks}  and graphs arise naturally in many complex systems, such as social networks, the World Wide Web, climate data, biological systems, and others. These networks normally encode relationships between two subjects within the system. For example, in the widely known Facebook friendship network, a connection is established between two users if they are friends on Facebook. Data of such type normally come in two versions: either a single static network that encodes all information at a particular time point, or a sequence of networks that captures the temporal evolution of the system. These networks can normally be classified into undirected and directed as well as weighted and unweighted networks. Unless stated otherwise, all networks described below are undirected unweighted networks. 

Analysis of static networks has been a popular research subject in both statistics and the social sciences. Of the many research topics one might pursue, community detection is arguably the most common choice. In brief, the goal of community detection is to locate highly dense sub-networks within the entire network \cite{newman04,fortunato10}. Among the many developed methods, modularity \cite{newman06} and statistical model based approaches \cite{holland83} have drawn much attention. Many algorithms have then been developed based on these ideas, including the famous Louvain method of \cite{louvain}, Infomap by \cite{infomap09}, and fast modularity by \cite{fast04}, among others \cite{huang2019, lyzinski2017}. A further important area of research in static network analysis is to determine the number of communities within a network \cite{airoldi08, Saldana15}.

While static network methods aim at analyzing individual snapshots of network data, time evolving network analysis tries to analyze a sequence of network data simultaneously.
There are two main approaches for community detection in time evolving networks:
consensus clustering, where one tries to find a community structure that fits well for all the snapshots in the data sequence, and change point detection, where one aims at locating the time points at which community structures change. 

In terms of consensus clustering, several major techniques have been developed in previous research \cite{multistep,consensus,ghase2016,wu2019,al2019}. 
These include sum graphs and average Louvain \cite{multistep}, which start by constructing a special graph that captures the topology of all snapshots in a given graph sequence and then apply any static community detection method to this summary graph. This assumes that the discovered structure fits well to all snapshots in the sequence. The construction of this special graph can be done in many ways, and the simplest way is to add up the adjacency matrices of each snapshot to create a new matrix that resembles this special graph (see Section~\ref{Optim} for more details). The method of \cite{consensus} aims to find a partition for the sequence using individual partitions from the snapshots. That is, using the community structure of each snapshot as input, the method constructs an adjacency matrix $M$ that captures the community assignment relationships between the nodes across all snapshots, and conducts community detection on $M$. Two algorithms were proposed by \cite{ghase2016} for detecting communities in time series network data under the assumption that the edges at each time point are independent. Since the algorithms are computationally expensive, the results from numerical simulations were only limited to the case of two communities.  A community detection algorithm was proposed by \cite{wu2019} for prediction using the spectral clustering. Lastly, \cite{al2019} combined spectral clustering and RPCA with constraints for smoothness on community evolution structure. 

As for change point detection, there have been a few well known methods that use different approaches on solving the problem. These include DeltaCon \cite{Koutra2016}, GraphScope \cite{graphscope}, Multi-Step \cite{multistep}, generalized hierarchical random graph (GHRG) \cite{GHRG}, and SCOUT \cite{SCOUT}. DeltaCon is a graph classification algorithm that compares graphs using the $L_2$ norm of the difference of the square-rooted adjacency matrices. GraphScope works by sequentially evaluating the next snapshot, to see whether its community structure matches well with the snapshot of the current segment according to an evaluation criterion derived from the Minimum Description Length Principle. This method also works well for online change point detection, i.e.~streaming data. However, the algorithm assumes the nodes can be partitioned into two sets: sources and sinks, and finds the partition within each set. Multi-Step, on the other hand, starts by assuming each snapshot belongs to its own segment. At each iteration, the two snapshots that are most similar (measured by an averaged modularity quantity) are grouped together. This is similar to a hierarchical clustering approach. GHRG works by first assuming a parametric model on the individual networks and a fixed-length moving window of the snapshots, and statistically tests whether a given time point in the window is a change point. Lastly, SCOUT works by finding the set of change points and community structures that minimizes an objective criterion derived from the Akaike Information Criterion (AIC) \cite{akaike} or Bayesian Information Criterion (BIC) \cite{BIC}. In \cite{SCOUT}, three change-point search algorithms are derived, respectively based on exhaustive search (with the use of dynamic programming to speed up computation), top-down search, and bottom-up search. Users can either pre-specify the number of change points and have the algorithm search over the restricted space, or let the algorithm determine the number of change points.

The proposed algorithm in this paper is designed to detect multiple change points and perform community detection simultaneously.  We note that \cite{marco2018} also developed an algorithm for doing the same task.  However, \cite{marco2018} assumed that the number and the composition of the communities do not change over time. Under this assumption, the edge behavior can be modeled by conditional non-homogeneous Poisson point processes.  The change points are estimated through the common discontinuity points in their intensity functions, assumed to be stepwise constant. On the other hand, the proposed algorithm allows the number of communities to change over time.  It is developed using a model selection framework in which the Minimum Description Length Principle (MDL) \cite{Rissanen89,Rissanen07} is utilized as minimizing objective criterion.  In other words, the proposed algorithm aims at selecting the number of change points and community assignments within each time segment by minimizing an objective criterion.
We note that, although GraphScope also uses the MDL principle as its objective criterion, model assumptions are different from the ones made in this paper. Unlike many of the existing papers, this paper provides a thorough analysis of the proposed method via simulated data, analyzing accuracy of the method when the ground truth is known --- an important validation step. It is specifically shown that, when the underlying model is correctly specified, the proposed method is able to detect change points with very high accuracy. Even when the model is misspecified (such as having correlated edges), the proposed method can still capture the change points, while competitor methods tend to over-estimate the locations in this scenario. The proposed algorithm is also designed to be more flexible in estimating certain parameters, which will be discussed in Section~\ref{MDL}.

While there are existing works that can perform change point detection and/or community detection in dynamic networks, there are certain limitations in most of these algorithms. For example, in \cite{peixoto15}, the models presented in the paper only concern the change in community assignments, while keeping the number of communities fixed and identical across different time segments. On the other hand, \cite{barucca18} and \cite{sewell17} considered a more flexible model by incorporating dependency between networks, which is a reasonable assumption in dynamic network modeling, but their model assumes a fixed number of communities across different segments as well. Lastly, \cite{deridder17} developed a more flexible algorithm that is built based on the work of \cite{GHRG}, by allowing a broader class of network models to be used in the algorithm. However, similar to \cite{GHRG}, the drawback of the algorithm is in requiring the user to choose the sliding window size for change point detection, with results potentially affected on the window size used. The proposed algorithm in this paper addresses most of these issues. While it does not consider temporal dependency, the added flexibility this model offers allows the user to gain a more fine-grained understanding of the data.

The rest of the paper is organized as follows. Section~\ref{problem.section} formally defines the problem. Sections~\ref{MDL} and~\ref{Optim} introduce the proposed methodology. Section~\ref{empirical} presents an empirical analysis of the proposed methodology and Section~\ref{conclude} concludes.

\section{Setting}
\label{problem.section}
\subsection{Notations}
\label{notations}
Denote a sequence of graphs of length $T$ as $\mathcal{G} = \{G^{(1)}, \dots, G^{(T)}\}$. Each graph $G^{(t)}$ consists of a vertex set $V^{(t)}$ and an edge set $E^{(t)}$, where the node degree of each $v \in V^{(t)}$ is at least 1. Note that there is no restriction on the size of the vertex sets: $|V^{(t)}|$ can be different from $|V^{(t^\prime)}|$ for $t \neq t^\prime$, implying that the graphs $G^{(t)}$ and $G^{(t^\prime)}$ have different size --- a quite natural assumption for time evolving networks. For example, in the popular Enron Email dataset \cite{enron}, each graph $G^{(t)}$ represents the email communication pattern between employees over one week. The nodes represent employees of the company, and an edge between two nodes means there is at least one email communication between the two employees within the time frame of the graph. It is possible that some employees have no email connection with the subjects of interest in the data set for some time $t$. Hence, these employees will be missing in $V^{(t)}$ but may be included at another time $t^\prime$ if email communication resumes. Denote the overall node set by $V = \bigcup_tV^{(t)}$, with $|V| = N$.

In general, each graph $G^{(t)} \in \mathcal{G}$ can be represented as a binary adjacency matrix $A^{(t)}$ of dimension $N \times N$, where $A^{(t)}_{ij} = 1$ represents a connection between the nodes $i$ and $j$, and 0 otherwise. If $|V^{(t)}| < N$, one can simply insert rows and columns of 0 at the appropriate locations such that the row and column arrangements of all matrices $A^{(t)}$ have the same meaning. Note that $\sum_jA_{ij}^{(t)} = 0$ means that no edge is connected to node $i$ (i.e.~node $i$ is a singleton). Of interest are the nodes such that $\sum_jA_{ij}^{(t)} \neq 0$, but for simplicity of notation and computation, all adjacency matrices are fixed at the same size. As stated in the Introduction, this paper focuses on simple undirected networks, hence $A^{(t)}_{ij} = A^{(t)}_{ji} = 1$ if there is at least one connection between node $i$ and $j$ at time $t$. The graph is also assumed to have no self-loops, i.e.~$A^{(t)}_{ii} = 0$. 

\subsection{Problem Statement}
\label{problem.statement}
Suppose the sequence of graphs $\mathcal{G}$ can be segmented into $M+1$ segments, with the graphs in each segment satisfying some homogeneity properties. For $m = 1, \dots, M+1$, define the graph segment $\mathcal{G}^{(m)} = \{G^{(t_{\{m-1\}})}, \dots, G^{(t_{m}-1)}\}$, using the conventions $t_0 = 1$ and $t_{M+1} = T+1$. The problem of change point detection in  networks can then be defined as follows:
\begin{problem}
\label{problem}
Given a sequence of graphs $\mathcal{G}$, find the locations $t_1,\ldots,t_M$ such that the community structure of each resulting graph segment $\mathcal{G}^{(m)}$ is homogeneous but different from the community structure of any adjoining graph segment.
\end{problem}
The time points $t_1,\ldots,t_M$ are called change point locations. It is important to note that, as mentioned above, the number of nodes within each graph can be different, even within the same time segment. However, if a change in community size is considered as a change in community structure, it can easily result in $T$ segments consisting of one graph each. Hence a more robust definition of `change' is needed in order to prevent overestimating the number of segments. 

\begin{definition}[Community structure within segment]
\label{assignment.def}
A community structure for segment $\mathcal{G}^{(m)}$ is a  partition of the node set $\mathcal{V}^{(m)} = \bigcup_{t_{\{m-1\}}}^{t_m-1}\tilde{V}^{(t)}$ into $c_m$ non-overlapping sets. The sets $\tilde{V}^{(t)}$ are called communities.
\end{definition}
It is possible that some nodes might not show up in all of the graphs in the segment. However, if there is evidence of a community structure (see Section~\ref{SBM}), adding nodes to the existing network can only strengthen the existing communities unless the new nodes introduce a large number of new connections. Similarly, removing certain nodes will not significantly weaken the existing structure unless the removed nodes play a central role in their communities. Hence this is a valid definition of community assignments. Because of this, for simplicity we will write $V^{(t)}$ instead of $\tilde{V}^{(t)}$ in what follows.

\section{Change Point Detection and Community Detection Using MDL}
\label{MDL}
This section describes the modeling procedure of the time evolving network and introduces the proposed methodology for change point and community detection. The statistical model used for the individual networks is presented first.

\subsection{The Stochastic Block Model}
\label{SBM}

Many statistical models have been proposed to analyze network data.  A popular example is the Stochastic Block Model (SBM). Below we briefly review the SBM.

Recall that the adjacency matrix is a symmetric binary matrix, with 1 representing the existence of a connection between two nodes. Given the community assignment vector $\mathbf{c}$ and link probabilities $P_{kl}$ between communities $k$ and $l$, one can model the edges with a Bernoulli distribution: $$A_{ij}|\mathbf{P}, \mathbf{c} \sim Ber(P_{c_ic_j}),$$ where $c_i$ and $c_j$ are the community assignments of nodes $i$ and $j$, and $\mathbf{P}$ is a symmetric matrix with $[\mathbf{P}]_{kl} = P_{kl}$. The standard assumption entails that $P_{kl}$ should be large if $k = l$. That is, if two nodes belong to the same community, there is a high probability of an edge existing between the two nodes. This results in a denser connection for intra-communities than for inter-communities. Extending this notation to the segmented setting mentioned above, we have $$A_{ij}^{(t)} | \mathbf{P}^{(t)}, \mathbf{c}^{(m)} \sim Ber(P^{(t)}_{c^{(m)}_ic^{(m)}_j})$$ if $G^{(t)}$ belongs to the $m^{th}$ segment. Note that the link probabilities $P_{kl}$ are not assumed to remain the same throughout a given segment, which allows the model to be more flexible and robust to minor changes in the network compositions.

The estimation of the link probabilities can be solved via the maximum likelihood method. Suppose the community assignment $\mathbf{c}$ at time $t$ is known, where $t \in [t_m, t_{m+1}-1]$. The log-likelihood function 
is then 
\begin{align}
&\mathcal{L}(\mathbf{P}^{(t)}|A^{(t)},\mathbf{c}^{(m)}) \nonumber \\
& = \sum_{i < j}[ A_{ij}^{(t)}\log(P^{(t)}_{c^{(m)}_ic^{(m)}_j}) + (1-A^{(t)}_{ij})\log(1-P^{(t)}_{c^{(m)}_ic^{(m)}_j})]
\nonumber \\
& = \sum_{k\leq l}[E^{(t)}_{kl}\log(P^{(t)}_{kl}) +
 (N^{(t)}_{kl}-E^{(t)}_{kl})\log(1-P^{(t)}_{kl})].
\label{berlikelihood2} 
\end{align}

The middle part of Equation~(\ref{berlikelihood2}) gives the representation when the edges are assumed to have Bernoulli distributions, while the right-hand side 
is for the aggregation of all edges within a given community into one group, with $1 \leq k \leq l \leq c_m$, using $N^{(t)}_{kl}$ as the total number of possible edges between communities $k$ and $l$, and $E_{kl}^{(t)}$ as the number of observed edges between communities $k$ and $l$. The parameters can then be estimated by finding the $\hat{P}^{(t)}_{kl}$ that maximize Equation~(\ref{berlikelihood2}).

\subsection{The MDL Principle}
Using the SBM as the base model for the graphs, one can write down a complete likelihood for modeling the change points and the community assignments for each segment (call this the segmented time-evolving network). As seen in Section~\ref{SBM}, the estimation of the link probabilities is trivial if the change point locations and community assignments are given. However, the estimation of the community structures and change points is less straightforward. In terms of community detection, various algorithms and objective criteria have been proposed to solve the problem (see Introduction). If the change point locations are known, one can easily adopt the existing methods to derive the community assignments. The rest of this section will apply the MDL principle to derive an estimate for the change point locations as well as community assignments for each segment.

The MDL principle is a model selection criterion. When applying the MDL principle, the ``best'' model is defined as the one allowing the greatest compression of the data $\mathcal{A} = (A^{(1)}, A^{(2)}, \dots, A^{(T)})$. That is, the ``best'' model enables us to store the data in a computer with the shortest code length. There are several versions of MDL, and the ``two-part'' variant will be used here (see \eqref{codelength} below).  The first part encodes the fitted model being considered, denoted by $\hat{\mathcal{F}}$, and the second part encodes the residuals left unexplained by the fitted model, denoted by $\hat{\mathcal{E}}$. Denote by $\text{CL}_{\mathcal{F}}(\mathcal{A})$ the code length of $\mathcal{A}$ under model $\mathcal{F}$, then 
\begin{align}
\label{codelength}
\text{CL}_{\mathcal{F}}(\mathcal{A}) = \text{CL}_{\mathcal{F}}(\hat{\mathcal{F}}) + \text{CL}_{\mathcal{F}}(\hat{\mathcal{E}}|\hat{\mathcal{F}}).
\end{align} 
The goal is to find the model $\hat{\mathcal{F}}$ that minimizes (\ref{codelength}). Readers can refer to \cite{Lee01} for more examples on how to apply the two-part MDL in different models. To use (\ref{codelength}) for finding the best segmentation as well as community assignments for a given evolving network sequence, the two terms on the right side of (\ref{codelength}) need to be calculated.

To fit a model for the segmented time-evolving network, one needs to first identify the change locations. Once the locations are determined, one can proceed to estimate the community assignments as well as the link probabilities. Denote by $\mathbf{c}^{(m)} = (c^{(m)}_1, ..., c^{(m)}_{|V^{(m)}|})$ the community assignment for the $m^{th}$ segment, and $\mathcal{C} = \{\mathbf{c}^{(1)}, \dots, \mathbf{c}^{(M+1)}\}$. Since $\hat{\mathcal{F}}$ is completely characterized by $\mathcal{T} = (t_1, \dots, t_M)$, $\mathcal{C}$, and $\mathcal{P} = \{\mathbf{P}^{(1)}, \dots, \mathbf{P}^{(T)}\}$, the code length of $\hat{\mathcal{F}}$ can be decomposed into
\begin{equation}
\label{modelcode}
\text{CL}_{\mathcal{F}}(\hat{\mathcal{F}}) = \text{CL}_{\mathcal{F}}(M) + \text{CL}_{\mathcal{F}}(\mathcal{T}) + \text{CL}_{\mathcal{F}}(\mathcal{C}) + \text{CL}_{\mathcal{F}}(\mathcal{P}).
\end{equation}
According to \cite{Rissanen89}, it requires approximately $\log_2I$ bits to encode an integer $I$ if the upper bound is unknown, and $\log_2I_u$ bits if $I$ is bounded from above by $I_u$. Hence $\text{CL}_{\mathcal{F}}(M)$, the code length for number of change points, translates to $\log_2(M+1)$, where the additional 1 is to differentiate between $M = 0$ (no change point) and $M = 1$. To encode the change point locations $\mathcal{T}$, one can encode the distances between each change point rather than the locations themselves. Hence $$\text{CL}_{\mathcal{F}}(\mathcal{T}) = \sum_{m = 1}^{M+1}\log_2(t_m - t_{\{m-1\}}+1).$$  Notice that another way to encode the change points is to encode them with code length $$\mbox{CL}'_{\mathcal{F}}(\mathcal{T})=M \log_2 T,$$ as all change point locations are upper-bounded by $T$.  This encoding method tends to encourage a set of more uniformly distributed change points. Comparing to $\mbox{CL}_{\mathcal{F}}(\mathcal{T})$, $\mbox{CL}'_{\mathcal{F}}(\mathcal{T})$ typically produces a larger code length.  As the MDL principle prefers coding methods that produce shorter code length (on average),  $\mbox{CL}_{\mathcal{F}}(\mathcal{T})$ is chosen to be in the procedure instead.  Also note that $\mbox{CL}_{\mathcal{F}}(\mathcal{T})$ and $\mbox{CL}'_{\mathcal{F}}(\mathcal{T})$ give very similar practical results, if not identical.

Once the change points are encoded, one can encode the community structures and link probabilities, i.e.~the networks themselves. Recall in Definition~\ref{assignment.def}, that the goal is to partition each node set $\mathcal{V}^{(m)}$ into $c_m$ non-overlapping communities. Therefore, $$\text{CL}_{\mathcal{F}}(\mathcal{C}) = \sum_{m = 1}^{M+1}\log_2c_m + |\mathcal{V}^{(m)}|\log_2c_m,$$ 
where the first term encodes the number of communities for the $m^{th}$ segment ($c_m \geq 1$), and the second term encodes the community assignment for each node. Lastly, by \cite{Rissanen89}, it takes $\frac12\log_2N$ bits to encode a maximum likelihood estimate of a parameter computed from $N$ observations. Hence, $$\text{CL}_{\mathcal{F}}(\mathcal{P}) = \sum_{t=1}^T\sum_{k \leq l}\frac{1}{2}\log_2N_{kl}^{(t)}.$$ Putting everything together,
\begin{align}
\label{modelcodelength}
\text{CL}_{\mathcal{F}}(\hat{\mathcal{F}}) =& \log_2(M+1)  + \sum_{m = 1}^{M+1}\log_2(t_m - t_{\{m-1\}}+1)\nonumber \\ 
& + \sum_{m = 1}^{M+1}\log_2c_m + |\mathcal{V}^{(m)}|\log_2c_m \nonumber \\
& + \sum_{t=1}^T\sum_{k \leq l}\frac{1}{2}\log_2N_{kl}^{(t)}.
\end{align}

To obtain the second term of (\ref{codelength}), one can use the result of \cite{Rissanen89} that the code length of the residuals $\hat{\mathcal{E}}$ is the negative of the log-likelihood of the fitted model $\hat{\mathcal{F}}$. With the assumption that, given the community structures and link probabilities, $A^{(t)}_{ij}$ follows a Bernoulli distribution, 
\begin{align}
\label{errorcodelength}
&\text{CL}_{\mathcal{F}}(\hat{\mathcal{E}}|\hat{\mathcal{F}})  \\
&=- \sum_{t=1}^T \sum_{k \leq l} [E_{kl}^{(t)}\log P_{kl}^{(t)} + (N_{kl}^{(t)}-E_{kl}^{(t)})\log(1 - P_{kl}^{(t)})]. \nonumber 
\end{align}

Combining (\ref{modelcodelength}) and (\ref{errorcodelength}) together, the proposed MDL criterion for estimating the change point locations and community structures is 
\begin{align}
\label{fullMDL}
\mbox{MDL}&(M, \mathcal{T},\mathcal{C}, \mathbf{P})= \log_2(M+1)  \\
+&\sum_{m = 1}^{M+1}\log_2(t_m - t_{\{m-1\}}+1) \nonumber \\
+& \sum_{m = 1}^{M+1}\log_2c_m + |\mathcal{V}^{(m)}|\log_2c_m \nonumber \\
+& \sum_{t=1}^T\sum_{k \leq l}\frac{1}{2}\log_2N_{kl}^{(t)}  \nonumber \\
-& \sum_{t=1}^T \sum_{k \leq l} [E_{kl}^{(t)}\log P_{kl}^{(t)} + (N_{kl}^{(t)}-E_{kl}^{(t)})\log(1 - P_{kl}^{(t)})].\nonumber
\end{align}
The goal is to find the change point locations and community assignments that minimize (\ref{fullMDL}). 

\section{Change Point and Community Assignment Search}
\label{Optim}
As pointed out in Section~\ref{SBM}, the estimates of link probabilities $\mathbf{P}$ are easy to obtain if the change points $\mathcal{T}$ and community assignments $\mathcal{C}$ are known. However, the estimation of $\mathcal{T}$ and $\mathcal{C}$ are non-trivial. Below describes the procedure for estimating these two parameters, which combine to estimate the segmented time-evolving network.

\subsection{Community Detection}
The procedure for community detection within a given segment of networks is described first. Recall that in Definition~\ref{assignment.def} the goal is to find, for the $m^{th}$ segment, $\mathbf{c}^{(m)}$ such that each node in $\mathcal{V}^{(m)} = \bigcup_{t = t_m}^{t_{m+1}-1}V^{(t)}$ belongs to exactly one community. However, it is possible that some nodes only appear in certain snapshots within the $m^{th}$ segment. Hence the community search procedure should be robust enough to deal with this problem.

Consider the set of adjacency matrices $\mathcal{A}^{(m)} = (A^{(t_m)}, \dots, A^{(t_{m+1}-1)})$. Suppose we can aggregate these matrices (which represent networks), by simply adding up these $t_{m+1} - t_m$ matrices. The resulting matrix forms a super network that overlays all the networks between $G^{(t_m)}$ and $G^{(t_{m+1}-1)}$, and community detection can be conducted over this super network. Since only simple undirected networks are considered, all values larger than 1 in the aggregated adjacency matrix will be replaced by 1. 

As seen in the Introduction, community detection has been a popular research area in the past few decades, and many known fast algorithms have been developed for the task. However, most of the designed algorithms aim at maximizing the modularity of the network, hence they cannot be applied directly here since the objective function of interest is the MDL criterion. Nonetheless, one can still borrow ideas from the algorithmic portion of the designed methodologies. 

The Louvain method of \cite{louvain} is known to be one of the fastest community detection algorithms for static networks. It works in the following way. First, all nodes are assigned to be their own communities. In the first iteration, each node (in some random order) is moved to its neighborhood community if there is a positive gain in modularity. If there are multiple neighborhood communities with positive gain, the one with maximum gain is picked. This is repeated for all nodes and perhaps multiple times per node until no modularity gain is achieved. Then the newly formed communities are treated as nodes and the merging procedure is repeated again until no modularity gain is achieved (at this step a neighborhood community is a group of vertices such that it has at least one connection with the current community). This method is fast and suitable for large graphs. However, in the statistical literature there is some evidence that bottom-up search methods tend to produce less optimal solutions; e.g., \cite{Lee02:genetic}.
Also, the number of communities is usually a lot smaller than the number of nodes, hence it seems not necessary to initialize $N$ communities with $N$ nodes in the graph.

Instead of a bottom-up search, a top-down algorithm for detecting communities is proposed here. The main idea is to recursively split the network into smaller communities until no further improvement can be achieved. The algorithm starts by randomly assigning each node to one of two communities. In the first iteration, each node (at some random order) is switched to the opposite community if the switch leads to a decrease in MDL value. Repeat this multiple times until no switch will cause a decrease in MDL value. Then, repeat the same procedure on each sub-community until no further split can be found. 

To prevent overestimating the number of communities, a merging step is conducted after the splits. At each iteration, each community is merged with its neighborhood community if there is a drop in MDL value, and the one with the biggest drop is picked if there are multiple such communities. Repeat this with all communities. One can think of this procedure as a top-down search (splitting communities) followed by a bottom-up search (merging communities). One can repeat the entire procedure after the merge step to prevent trapping in a locally optimal solution. 

Notice that since all the segments are assumed to be independent of each other, there is no need to calculate the entire MDL value~(\ref{fullMDL}) when conducting community search. Instead, one can consider the sub-MDL criterion
\begin{eqnarray}
\label{subMDL}
&&\log_2c_m + |\mathcal{V}^{(m)}|\log_2c_m + \sum_{t=t_m}^{t_{m+1}-1}\sum_{k \leq l}\frac{1}{2}\log_2N_{kl}^{(t)} \\
&&- \sum_{t=t_m}^{t_{m+1}-1} \sum_{k \leq l} [E_{kl}^{(t)}\log P_{kl}^{(t)} + (N_{kl}^{(t)}-E_{kl}^{(t)})\log(1 - P_{kl}^{(t)})] \nonumber
\end{eqnarray}
when doing the splitting and merging steps mentioned in the previous paragraphs. This also means that all segments can be searched simultaneously, which then speeds up computational time. Algorithm~\ref{alg:cd} lays out the community assignments search procedure.

\begin{algorithm}
\small
\caption{Community Detection for the $m^{th}$ Segment}
\label{alg:cd}
\begin{algorithmic}[1]
\STATE Assign each node to one of two communities. To speed up the initialization process, use existing methods to identify the two communities.
\STATE Calculate the MDL value using (\ref{subMDL}). Denote thus value by $\mathrm{MDL}_m$.
\WHILE{there is a drop in $\mathrm{MDL}_m$}
	\FOR{each node in $\mathcal{V}^{(m)}$}
		\STATE Switch the community assignment if the value of (\ref{subMDL}) is lowered. Update $\mathrm{MDL}_m$.
	\ENDFOR
\ENDWHILE
\IF{there is no community found} 
	\STATE Stop.
\ENDIF
\WHILE{there is a drop in $\mathrm{MDL}_m$}
	\FOR{each community found}
		\STATE Repeat steps 1-6, but with a subset of $\mathcal{V}^{(m)}$.
	\ENDFOR
	\STATE Update $\mathrm{MDL}_m$.
\ENDWHILE
\STATE Merge communities until no drop in $\mathrm{MDL}_m$.
\STATE Repeat steps 10-17 until no drop in $\mathrm{MDL}_m$.
\STATE Return the community assignments $\mathbf{c}^{(m)}$.
\end{algorithmic}
\end{algorithm}

\subsection{Change Point Detection}
Change point detection algorithms for networks usually involve a top-down search, bottom-up search, or exhaustive search. An exhaustive search aims at finding the set of change point locations that minimizes the objective criterion by enumerating all possible combinations of change points. By doing so, the solution is guaranteed to be a global minimum, but the computation also becomes intractable once $T$ is large (with $T$ snapshots, there are $2^{T-1}$ combinations to loop through). One can use dynamic programming to reduce the computational complexity, but still needs to search through a large solution space before finalizing a global solution.

Both top-down and bottom-up searches are greedy algorithms, and their computation can be complex. For top-down search, one starts with the entire sequence of $T$ graphs, and finds the location $t_1 \in [2,T]$ that minimizes the objective criterion (as well as a decrease in the criterion value). Then, one finds the location $t_2 \in [2,T]\backslash {t_1}$ that minimizes the objective criterion (with $t_1$ already in the model), and repeats until no change point can be found. By doing so, one needs to go through $T-i$ calculations at the $i^{th}$ iteration. Bottom-up search, on the other hand, starts by assuming each location $t$ is a change point, and merge the adjacent segment such that the objective function is minimized. This procedure is repeated until no further merge can be found.

This paper proposes a top-down search for finding the change point locations. However, instead of naively testing each location for the possibility of being a change location, a screening process is first conducted to select a set of candidate change locations. Then each candidate location (in some specific order) is checked to see whether it is a change point or not. The details of the search algorithm are described below.

The screening process is conducted as follows. First, calculate the difference between each pair of consecutive adjacency matrices. The distance used is the 1-norm between the two matrices normalized by their geometric means, given by the formula

\begin{eqnarray}
\label{distance}
d_t = d(A^{(t-1)}, A^{(t)}) = \frac{||\mathrm{vec}(A^{(t-1)}) -  \mathrm{vec}(A^{(t)})||_1}{\sqrt{||\mathrm{vec}(A^{(t-1)})||_1||\mathrm{vec}(A^{(t)})||_1}}, 
\end{eqnarray}
where $\mathrm{vec}(A)$ is the vector form of $A$. The idea is that if the community structure between two consecutive networks does not change, then regardless of the differences in link probabilities, the edge pattern should remain roughly the same, hence the distance should be relatively smaller. Therefore, a large value of $d_t$ is an indicator that there is a change in the community structure at time $t$. Set the locations whose distances are above the median value of $d_t$'s as the candidate change locations. This is equivalent to assuming that the maximum number of change points is $T/2$, which is a reasonable assumption in most situations.

Once the candidate locations are determined, order them by the $d_t$ values from largest to smallest. Starting with the first location (denoted by $\hat{t}_1$), segment the data into two pieces, conduct community search within each segment, and calculate the MDL value (\ref{fullMDL}). If this value is smaller than the MDL value with no segmentation, set $\hat{t}_1$ as a change location, otherwise segment the data at $\hat{t}_2$ and repeat. Every time a change location is found, remove the location from the candidate set and reset the search procedure, with the previously selected locations in the estimated model. Doing this requires at most $[(1+T/2)T/2]/{2} = [{2T + T^2}]/{8}$ calculations. Even though this can be large if $T$ is large, often times the search procedure will stop after a few iterations. 

To prevent overestimating the number of change points, a merging step is conducted on the selected change points (if any). There are two cases to consider: (1) at least one change point is selected in the previous step and (2) no change point is selected in the previous step. For case 
(1), merge the segments at the selected change locations, starting from the last selected change point, and recalculate the MDL value. If there is a decrease in the MDL value, keep the merge, otherwise ignore it, and move onto the next selected change point, until all estimated change points have been tested. For case (2), use the candidate locations (in reversed order $\hat{t}_{T/2}, \dots, \hat{t}_1$) as estimated change points, and perform the merging step. One can view this as a bottom-up search strategy. Algorithm~\ref{alg:cp} lays out the change points search procedure.

\begin{algorithm}
\small
\caption{Change Point Detection in time evolving Networks.}
\label{alg:cp}
\begin{algorithmic}[1]
\STATE Calculate the consecutive distances $d_t$ using (\ref{distance}) for adjacency matrices $A^{(2)}, \dots, A^{(T)}$.
\STATE Set $\tau_c = \{t\colon d_t \geq med(d_2, \dots, d_T)\}$. Order $\tau_c$ according to the values of selected $d_t$ from largest to smallest. Set $\tau = \emptyset$.
\STATE Calculate the MDL value (\ref{fullMDL}) for when there is no change point. Denote as $\mathrm{MDL}_{\rm old}$.
\FOR{each $t \in \tau_c$}
	\STATE Segment the network sequence at time $t$ (given change points at $t' \in \tau$) and conduct community detection with Algorithm~\ref{alg:cd}. 
	\STATE Calculate the MDL value (\ref{fullMDL}) using the segmented model. Denote as $\mathrm{MDL}_{\rm new}$.
	\IF{$\mathrm{MDL}_{\rm new} < \mathrm{MDL}_{\rm old}$} 
		\STATE $\tau = \tau \cup {t}$, $\tau_c = \tau_c \backslash \{t\}$, $\mathrm{MDL}_{\rm old} = \mathrm{MDL}_{\rm new}$. Restart for loop.
	\ENDIF
\ENDFOR
\IF{$\tau = \emptyset$}
	\STATE Set $\tau = \mathrm{rev}(\tau_c)$. Update $\mathrm{MDL}_{\rm old}$ using $\tau$ as change points.
\ELSE
	\STATE Set $\tau = \mathrm{rev}(\tau)$.
\ENDIF
\FOR{each $t \in \tau$}
	\STATE Merge the consecutive segments at $t$ and conduct community detection with Algorithm~\ref{alg:cd} (given change points at $t' \in \tau$). 
	\STATE Calculate the MDL value (\ref{fullMDL}) using the segmented model. Denote as $\mathrm{MDL}_{\rm new}$.
	\IF{$\mathrm{MDL}_{\rm new} < \mathrm{MDL}_{\rm old}$} 
		\STATE $\tau = \tau \backslash \{t\}$, $\mathrm{MDL}_{\rm old} = \mathrm{MDL}_{\rm new}$. Restart for loop.
	\ENDIF
\ENDFOR
\STATE Return $\tau$ and community structures with $\tau$ as change points.
\end{algorithmic}
\end{algorithm}

\section{Empirical Analysis}
\label{empirical}
To assess the performance of the proposed methodology, multiple simulation sets were conducted. Application to a data set was also performed to showcase the practical use of the proposed method. 

\subsection{Simulation}
This section focuses on analyzing the performance of the proposed method on synthetic data. Out of the six settings compared, five settings involved networks generated according to the SBM discussed in Section~\ref{SBM}, with each time shot independent of each other. The last setting involved networks with correlated edges, which were studied by \cite{Saldana15}. Change point detection results were compared with the Multi-Step change point detection algorithm of \cite{multistep}, and the SCOUT algorithm of \cite{SCOUT}. Publicly available implementations of both algorithms were used. Table~\ref{simulation} shows a summary of each setting. Detailed descriptions of each setting can be found in the appendix. Figures~\ref{setting1.result}-\ref{setting6.result} show the histograms of the estimated change point locations for Settings 1 through 6, respectively. All settings were repeated 100 times. 

\begin{table*}
\centering
\caption{Simulation settings. Detailed descriptions can be found in the appendix. ($\rho$ is the correlation between edges. See \cite{Saldana15}.)}
\label{simulation}
\begin{tabular}{|c|l|l|l|c|p{6cm}|}
\hline
Setting & Correlated Edges & Sparse/Dense & \# of Change Points & \# of Nodes Per Network & Remarks \\
\hline
1 & No & Dense & 5 & 280 - 300 & Networks within the same segment have same edge probabilities\\
2 & No & Dense & 4 & 280 - 300 & Each graph has different probability \\
3 & No & Moderate & 4 & 380 - 400 & Each graph has different probability \\
4 & No & Sparse & 5 & 380 - 400 & Network within the same segment have same edge probabilities \\
5 & No & Sparse & 4 & 380 - 400 & Each graph has different probability \\
6 & Yes & Dense & 5 & 380 - 400 & $\rho = 0.7$\\
\hline
\end{tabular}
\end{table*}

\begin{figure}[htbp]
\centering
\includegraphics[width = 0.5\textwidth, height = 0.2\textheight]{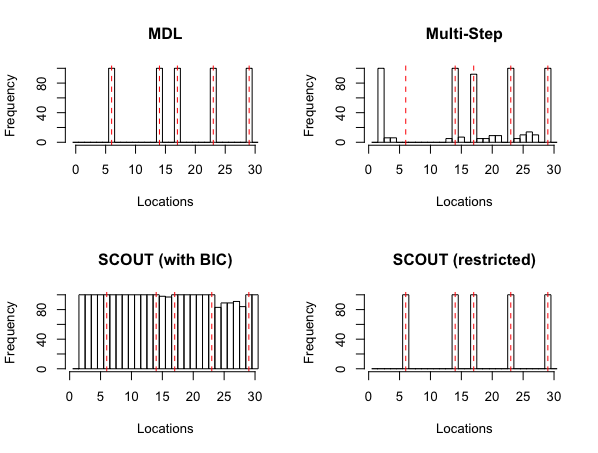}
\caption{The red dotted vertical lines represent the true change points. The heights of the bars represent the frequency of the estimated change points with simulation setting 1 over 100 trials. For SCOUT (with BIC) - using BIC to select the number of change points; (Restricted) - restricting to the known number of change points.}
\label{setting1.result}
\end{figure}

\begin{figure}[htbp]
\centering
\includegraphics[width = 0.5\textwidth, height = 0.2\textheight]{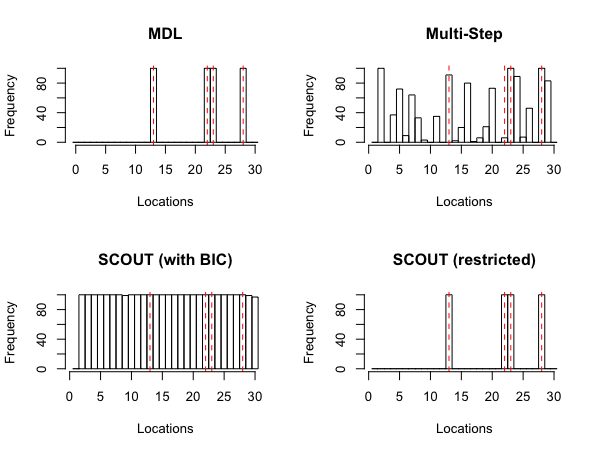}
\caption{The red dotted vertical lines represent the true change points. The heights of the bars represent the frequency of the estimated change points with simulation setting 2 over 100 trials. For SCOUT (with BIC) - using BIC to select the number of change points; (Restricted) - restricting to the known number of change points.}
\label{setting2.result}
\end{figure}

\begin{figure}[htbp]
\centering
\includegraphics[width = 0.5\textwidth, height = 0.2\textheight]{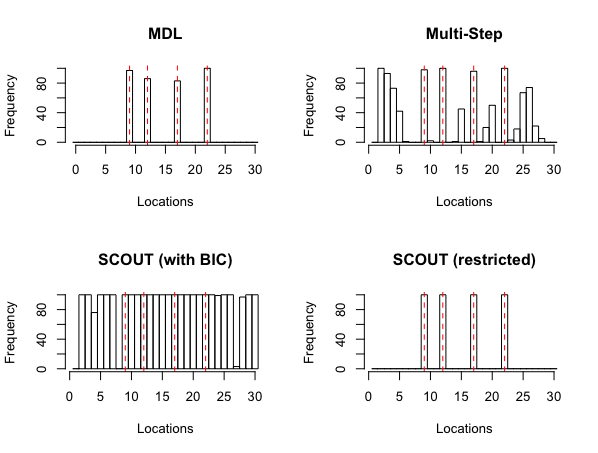}
\caption{The red dotted vertical lines represent the true change points. The heights of the bars represent the frequency of the estimated change points with simulation setting 3 over 100 trials. For SCOUT (with BIC) - using BIC to select the number of change points; (Restricted) - restricting to the known number of change points.}
\label{setting3.result}
\end{figure}

\begin{figure}[htbp]
\centering
\includegraphics[width = 0.5\textwidth, height = 0.2\textheight]{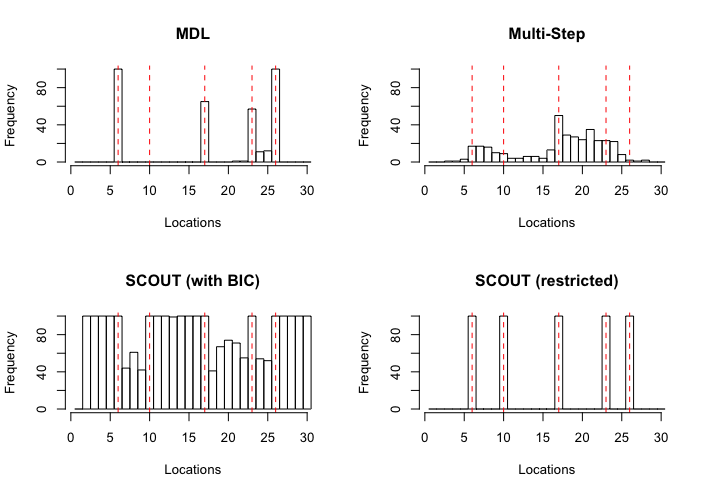}
\caption{The red dotted vertical lines represent the true change points. The heights of the bars represent the frequency of the estimated change points with simulation setting 4 over 100 trials. For SCOUT (with BIC) - using BIC to select the number of change points; (Restricted) - restricting to the known number of change points.}
\label{setting4.result}
\end{figure}

\begin{figure}[htbp]
\centering
\includegraphics[width = 0.5\textwidth, height = 0.2\textheight]{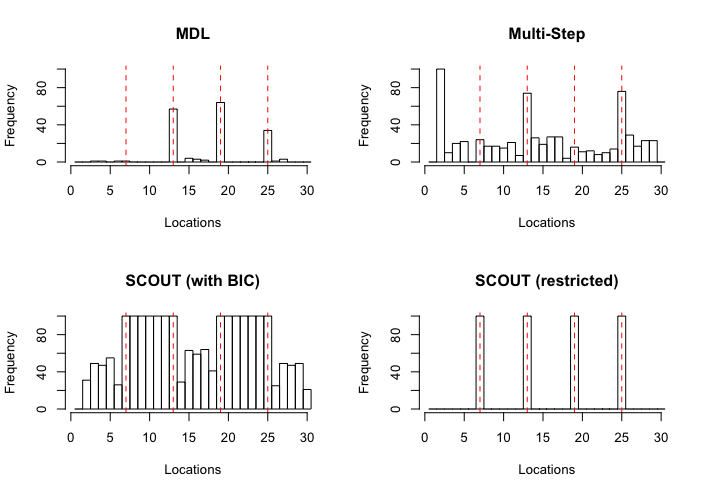}
\caption{The red dotted vertical lines represent the true change points. The heights of the bars represent the frequency of the estimated change points with simulation setting 5 over 100 trials. For SCOUT (with BIC) - using BIC to select the number of change points; (Restricted) - restricting to the known number of change points.}
\label{setting5.result}
\end{figure}

\begin{figure}[htbp]
\centering
\includegraphics[width = 0.5\textwidth, height = 0.2\textheight]{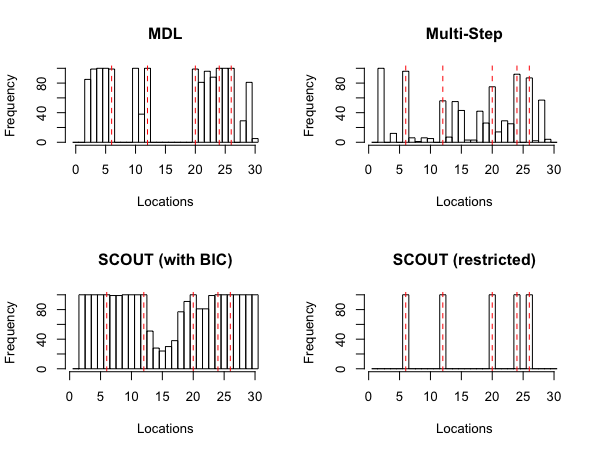}
\caption{The red dotted vertical lines represent the true change points. The heights of the bars represent the frequency of the estimated change points with simulation setting 6 over 100 trials. For SCOUT (with BIC) - using BIC to select the number of change points; (Restricted) - restricting to the known number of change points.}
\label{setting6.result}
\end{figure}

As listed in Table~\ref{simulation}, four of the settings involved dense networks while the remaining two involved sparse networks. A network is considered dense if the intra community edge probabilities is at least 0.4 while a network is considered sparse otherwise. Networks within the same segment in Settings 1, 4 and 6 have the same link probabilities, while the networks in the other settings have the different link probabilities even for networks within the same segment. Also, the intra probability links are set to be closer to the inter probability links in Settings 4 and 5, to demonstrate the scenarios where there is not a clear separation of communities.

From Figures~\ref{setting1.result}--\ref{setting3.result}, one can see that the proposed MDL method outperforms Multi-step in almost all cases. In Settings 4 and 5, while the proposed method fails to detect one of the change points in each setting, it can consistently locate the other change points, whereas Multi-step also estimates many spurious change points.
Lastly, even though the proposed method overestimated the number of change points in Setting 6, Multi-step seemed to be consistently overestimating as well. As for the SCOUT method, even though it always estimates the correct change point locations when the number of change points is known, it is seen that the method was not able to correctly identify the number of change points once this restriction was relaxed. As the number of change points is often unknown in real data, it is more reasonable to compare with results of the automatic selection case (with BIC). 

To also evaluate the performance of the proposed community detection algorithm, the normalized mutual information (NMI) was used. In brief, NMI is an evaluation criterion used to evaluate the performance of clustering results. It is defined as
\begin{eqnarray}
\label{NMI}
\mathrm{NMI}(\hat{\mathbf{c}}^{(m)}, \mathbf{c}^{(m)}) &=& \frac{I(\hat{\mathbf{c}}^{(m)}, \mathbf{c}^{(m)})}{[H(\hat{\mathbf{c}}^{(m)}) + H(\mathbf{c}^{(m)})]/2}.
\end{eqnarray}
Denote 
\[\mathcal{V}^{(m)}_l = \{v \in \mathcal{V}^{(m)} \colon v \text{ belongs to community } l\}\] and \[\mathcal{\hat{V}}^{(m)}_l = \{v \in \mathcal{V}^{(m)} \colon v \text{ is estimated to community } l\}.\] The quantities $H(\cdot)$ (entropy) and $I(\cdot)$ (mutual information) are then defined as
\begin{align}
\label{entropy}
H(\mathbf{c}^{(m)}) =& -\sum_lP(\mathcal{V}^{(m)}_l)\log P(\mathcal{V}^{(m)}_l)\nonumber \\
=& -\sum_l\frac{|\mathcal{V}^{(m)}_l|}{|\mathcal{V}^{(m)}|}\log\frac{|\mathcal{V}^{(m)}_l|}{|\mathcal{V}^{(m)}|}
\end{align}
and
\begin{align}
\label{MI}
I&(\hat{\mathbf{c}}^{(m)}, \mathbf{c}^{(m)}) \nonumber \\ 
&= \sum_k\sum_lP(\mathcal{\hat{V}}^{(m)}_k \cap \mathcal{V}^{(m)}_l) \log \frac{P(\mathcal{\hat{V}}^{(m)}_k \cap \mathcal{V}^{(m)}_l)}{P(\mathcal{\hat{V}}^{(m)}_k)P(\mathcal{V}^{(m)}_l)} \nonumber \\
&= \sum_k\sum_l\frac{|\mathcal{\hat{V}}^{(m)}_k \cap \mathcal{V}^{(m)}_l|}{|\mathcal{V}^{(m)}|}\log \frac{|\mathcal{V}^{(m)}||\mathcal{\hat{V}}^{(m)}_k \cap \mathcal{V}^{(m)}_l|}{|\mathcal{\hat{V}}^{(m)}_k||\mathcal{V}^{(m)}_l|}.
\end{align}
The overall NMI for the sequence of networks is defined as the mean of all individual NMIs: $\mathrm{NMI}_{\rm overall} = \frac{1}{M+1}\sum_m \mathrm{NMI}(\hat{\mathbf{c}}^{(m)}, \mathbf{c}^{(m)})$. Notice that $\mathrm{NMI}(\hat{\mathbf{c}}^{(m)}, \mathbf{c}^{(m)})$ ranges between 0 and 1, where 0 means the estimated community structure is a complete random guess, while 1 means it is perfectly matched. Table~\ref{comm.result} presents the results for community detection of the proposed algorithm, as well as detection results from SCOUT.  
As this set of comparison is to verify the performance of community detection of the proposed method and will not work well if the number of detected change points are different, only the case where the change points are known is considered.


\begin{table}[htbp]
\centering
\caption{Community detection results. Results show averages over 100 trials.}
\label{comm.result}
\begin{tabular}{|c|cccccc|}
\hline
Settings & 1 & 2 & 3 & 4 & 5 & 6\\ \hline
MDL & 1.00 & 1.00 & 0.55 & 0.527 & 0.227 & 0.83\\
SCOUT & 0.09 & 0.11 & 0.10 & 0.015 & 0.013 & 0.10\\
\hline
\end{tabular}
\end{table}

\subsection{Data Analysis}
In this application, the World Trade Web (WTW), also known as the International Trade Network (ITN), is considered. This data set is publicly available at \cite{WTW}. In brief, this data set captures the trading flow between 196 countries from 1948 to 2000. The data set consists of the total amount of imports and exports between two countries each year. Several papers have been published on the analyses of the WTW, including \cite{tdr08}, \cite{bmm}, \cite{bmsm} and \cite{bfm}. Since the import/export information is given, many of these analyses were done considering the trade network as a directed weighted network, where the weights represent the amount of goods going from country A to country B. For this analysis, however, since the focus of this paper is on undirected networks, the data set was modified such that an edge exists between two countries if there is some trading between the two countries. As there is data for each year between 1948 to 2000 (53 years), it is straightforward to consider this as a time evolving network. It should be noted that there can be different ways to create the adjacency matrices, such as having an edge between two countries only if the trade volume exceeds a certain threshold. However, empirical analyses show that the resulting change points are similar regardless of the threshold value used, hence the results are omitted here.
Table~\ref{data.detail} shows the summary of the data set.
\begin{table}[htbp]
\centering
\caption{Summary of data set}
\label{data.detail}
\begin{tabular}{cccc}
\hline
\# of nodes & \# of edges (mean $\pm$ SD) & Time span & Duration \\
\hline
196 & 5736 $\pm$ 2804 & 53 years & 1 year \\
\hline
\end{tabular}
\end{table}

\begin{figure}[htbp]
\centering
\includegraphics[width = 0.5\textwidth, height = 0.2\textheight]{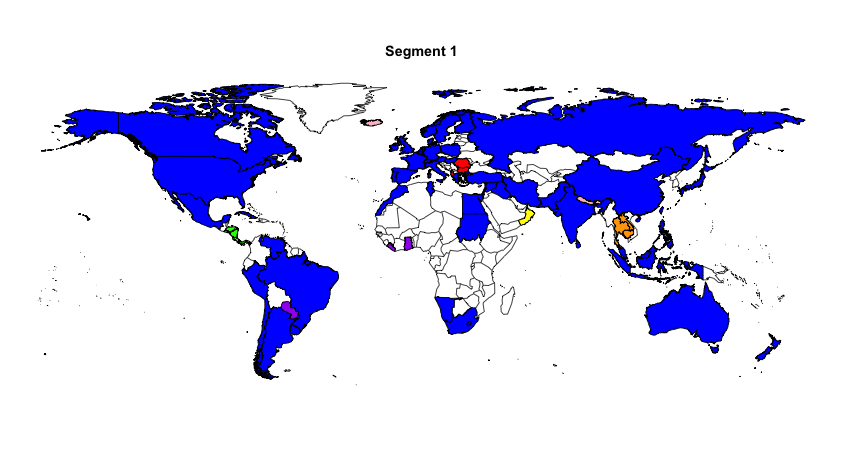}
\caption{Segment 1: 1948 - 1959. Largest 7 communities detected using proposed algorithm. Blue denotes the largest, green the second largest, then yellow, red, pink, orange, and purple for the third to seventh largest communities, respectively.}
\label{trade.1}
\end{figure}

\begin{figure}[htbp]
\centering
\includegraphics[width = 0.5\textwidth, height = 0.2\textheight]{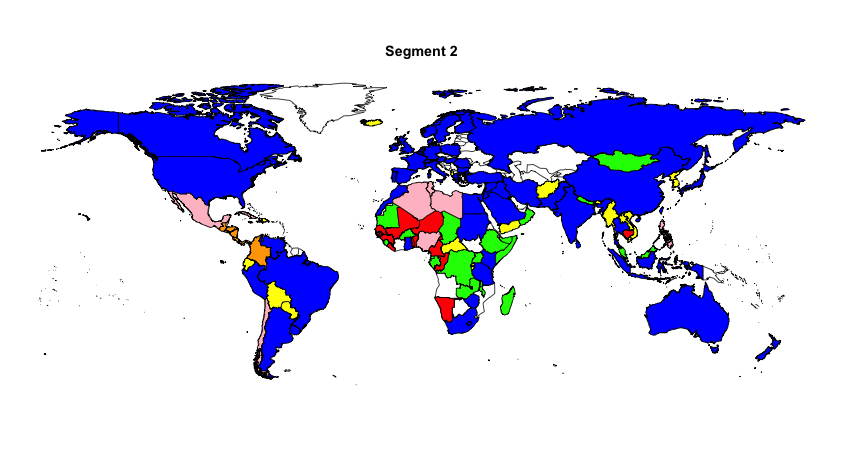}
\caption{Segment 2: 1960 - 1965.}
\label{trade.2}
\end{figure}

\begin{figure}[htbp]
\centering
\includegraphics[width = 0.5\textwidth, height = 0.2\textheight]{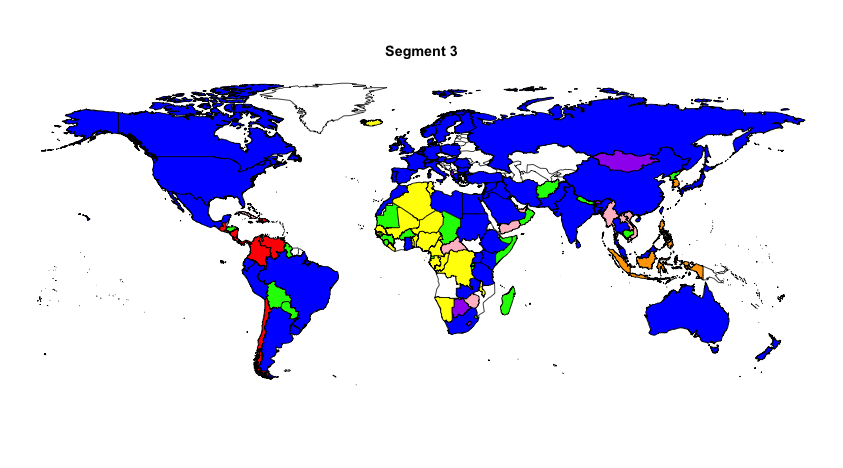}
\caption{Segment 3: 1966 - 1974.}
\label{trade.3}
\end{figure}

\begin{figure}[htbp]
\centering
\includegraphics[width = 0.5\textwidth, height = 0.2\textheight]{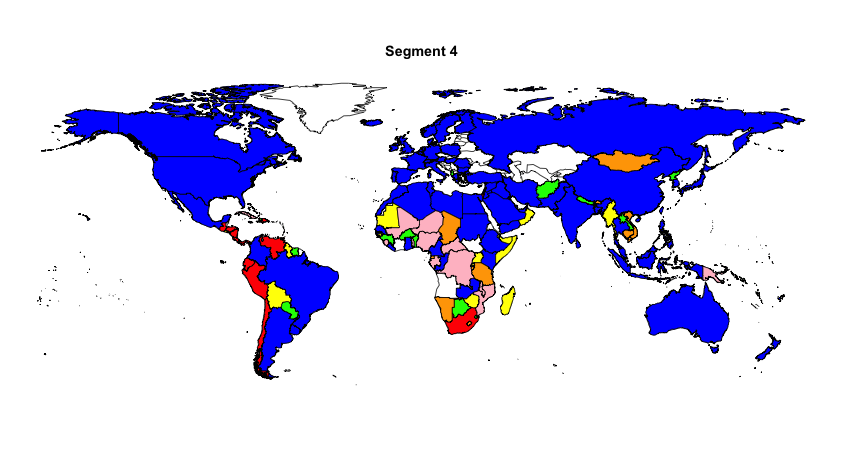}
\caption{Segment 4: 1975 - 1980.}
\label{trade.4}
\end{figure}

\begin{figure}[htbp]
\centering
\includegraphics[width = 0.5\textwidth, height = 0.2\textheight]{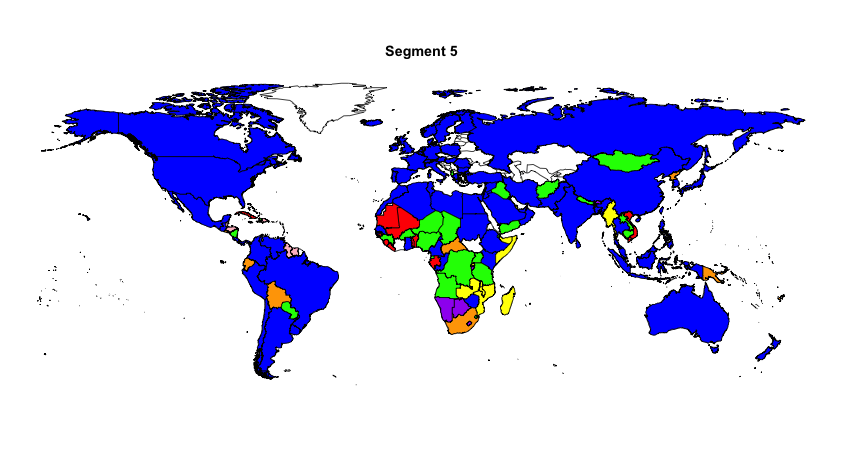}
\caption{Segment 5: 1981 - 1990.}
\label{trade.5}
\end{figure}

\begin{figure}[htbp]
\centering
\includegraphics[width = 0.5\textwidth, height = 0.2\textheight]{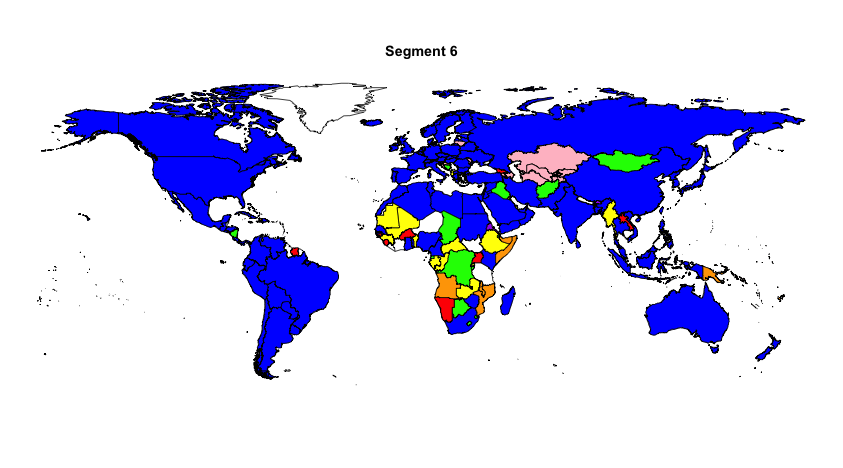}
\caption{Segment 6: 1991 - 2000.}
\label{trade.6}
\end{figure}

The proposed algorithm detected 5 change points on this data set. As comparison, the SCOUT algorithm (with BIC to select the number of change points) was also applied to the data set, which detected 5 change points as well. The results are listed in Table~\ref{data.result} below. 
\begin{table*}[htbp]
\centering
\caption{Segments determined by proposed algorithm and SCOUT.}
\label{data.result}
\begin{tabular}{c|cccccc}
\hline
& Segment 1 & Segment 2 & Segment 3 & Segment 4 & Segment 5 & Segment 6\\
\hline
MDL & 1948 - 1959 & 1960 - 1965 & 1966 - 1974 & 1975 - 1980 & 1981 - 1990 & 1991 - 2000\\
SCOUT & 1948 - 1961 & 1962 - 1972 & 1973 - 1980 & 1981 - 1990 & 1991 - 1992 & 1993 - 2000 \\
\hline
\end{tabular}
\end{table*}
Only the community assignments of the proposed method will be investigated here. Figures~\ref{trade.1}-\ref{trade.6} show the trading communities for the six detected segments. Since multiple communities have been detected for each segment, only the top 7 largest communities will be analyzed for each time period (the top 7 communities cover a majority of the countries in most cases). For each map, blue denotes the largest community, green the second largest, then yellow, red, pink, orange, and purple, for the third to seventh largest communities, respectively. One can see that the largest community consists of all the largest nations in the world, including the US, Canada, China, Russia, and many others.

The change points located by the two algorithms are similar, hence the following analysis will focus on change points detected by the proposed method. First notice that, as a consequence of our take on the data, changes in the network structure appear mostly for developing countries as the economies of developed countries are very robust even under major historical events, leading to a large community comprising all major economies over the entire observation period. As time progresses and their economies develop, further countries are added to this community, while occasionally countries drop out too. The majority of the changes in the network are confined to Africa, with some notable exceptions to be discussed below. 

During the first segment (1948 to 1959), large parts of Africa were still under colonial rule and do not appear in trade counts. 
Starting from the second segment (1960 to 1965), after their independence from European colonial powers most countries in Africa started to get involved in trading, forming their own communities. 
During the third segment (1966 to 1974), most trade behaviors in Africa remained stable. An interesting phenomenon is observed in South East Asia, where Indonesia broke off from the large community and formed a new group with several other countries. During this period, Indonesia transitioned from a communist regime to an autocratic regime supported by the military, leading to mass killings in 1965 and 1966. 
Only minor adjustments of the communities occurred in the fourth segment (1975 to 1980), and, in particular, Indonesia rejoined the large community. 
Throughout the last two segments, as a consequence of expanded worldwide economic integration (``globalization''), countries in Africa and South America started to join the large trade network with the biggest world economies. In the sixth segment (1991 to 2000), almost all countries in South America had joined (except for Suriname).

\section{Conclusion}
\label{conclude}
This paper presented a novel methodology for analyzing time evolving network data. By assuming each individual network follows a Stochastic Block Model (SBM), an objective criterion based on the Minimum Description Length (MDL) Principle was derived for detecting change points and community structures within each time segments in time evolving networks. In numerical simulation, our method shows promising results compared to the competitor methods Multi-Step and SCOUT. In addition, a data analysis revealed the proposed methodology is able to detect major changes.

In the above the dependence amongst networks was not taken into account.  Such an assumption of independence can be restrictive for various real life problems.  To relax this, one can adopt a Markov structure as in \cite{sewell17}.  For example, one can model the link probability $P_{ij}(t)$ at time $t$ between nodes $i$ and $j$ as
$$
P_{ij}(t) = 
\begin{cases}
p &\mbox{if nodes $i$ and $j$ are linked at time $t-1$}\\
1-p & \mbox{otherwise}
\end{cases},
$$
where $0\leq p \leq 1$ and its value varies in different segments.  An MDL criterion can be derived for this Markov modeling; e.g., \cite{Craiu-Lee06} applied MDL to a Markov modeling problem in image anaylsis.  The challenge is how to minimize such a criterion.  This will be investigated as future work.


%



\appendix[Simulation Details]
This appendix provides the details of the simulation settings.

\noindent \textbf{Setting 1:} Table \ref{Setting1} lists the specification for this setting. $T = 30$ for this and all following settings. The number of nodes for each snapshot ranged between 280 to 300. The community sizes were specified according to the ratios listed in the column `Community Size Ratio': the ratios (1/3, 1/3, 1/3) mean there are three communities, each containing roughly 1/3 of the total nodes of the graph. The link probabilities are listed in the `Link Probability' column, with $P_W$ representing the probability of an edge existing within a community, and $P_B$ the probability of an edge existing between two communities. Note the quantities satisfy the assumption $P_W > P_B$. For this setting, all networks within the same segment had the same within and between links probabilities. The true segments are listed in the column `Segment Number'. 
\begin{table}[htbp]
\tiny
\centering
\caption{Specifications for Setting 1.}
\label{Setting1}
\begin{tabular}{c|c|c|c|c}
\hline
Segment & $t$ & Community Size Ratio & Link Probability & \# of Nodes\\
\hline
1 & 1 - 5 & 1/3, 1/3, 1/3 & $P_W = 0.90, P_B = 0.10$ & 280 - 300 \\
2 & 6 - 13 & 1 & $P_W = 0.70, P_B = 0.20$ & 280 - 300 \\
3 & 14 - 16 & 1/4, 1/4, 1/4, 1/4 & $P_W = 0.85, P_B = 0.15$ & 280 - 300 \\
4 & 17 - 22 & 2/3, 1/3 & $P_W = 0.84, P_B = 0.20$ & 280 - 300 \\
5 & 23 - 28 & 1/5, 1/5, 1/10, 3/10, 1/5 & $P_W = 0.80, P_B = 0.15$ & 280 - 300 \\
6 & 29 - 30 & 3/10, 2/5, 3/10 & $P_W = 0.90, P_B = 0.10$ & 280 - 300\\
\hline
\end{tabular}
\end{table}

\noindent \textbf{Setting 2:} The previous setting assumed the link probabilities remain the same within each segment. However, this is not necessarily a valid assumption in real world data. This setting provides a setup such that each graph has a different intra and inter link probability. For all graphs, the intra and inter-link probabilities followed Uniform distributions: $P_W \sim U(0.70, 0.95)$ and $P_B \sim U(0.05, 0.3)$. The rest of the specifications are listed in Table \ref{Setting2}.

\begin{table}[htbp]
\centering
\caption{Specifications for Setting 2.}
\label{Setting2}
\begin{tabular}{c|c|c|c}
\hline
Segment & $t$ & Community Size Ratio & \# of Nodes\\
\hline
1 & 1 - 12 & 1/3, 1/3, 1/3 & 280 - 300 \\
2 & 13 - 21 & 1/3, 2/3 & 280 - 300 \\
3 & 22 - 22 & 3/4, 1/4 &  280 - 300 \\
4 & 23 - 27 & 3/10, 2/5, 3/10 & 280 - 300 \\
5 & 28 - 30 & 1/5, 3/10, 1/5, 3/10 & 280 - 300 \\
\hline
\end{tabular}
\end{table}

\noindent \textbf{Setting 3:} Both settings considered so far consist of dense networks. Often times, however, observed networks have a moderately dense structure. Instead of having a high $P_W$ value, this setting used $P_W \sim U(0.35, 0.40)$ and $P_B \sim U(0.05, 0.10)$. The rest of the specifications are listed in Table~\ref{Setting3}.

\begin{table}[htbp]
\centering
\caption{Specifications for Setting 3.}
\label{Setting3}
\begin{tabular}{c|c|c|c}
\hline
Segment & $t$ & Community Size Ratio & \# of Nodes\\
\hline
1 & 1 - 8 & 1/3, 1/3, 1/3 & 380 - 400 \\
2 & 9 - 11 & 1/4, 3/4 & 380 - 400 \\
3 & 12 - 16 & 1/2, 1/2 &  380 - 400 \\
4 & 17 - 21 & 3/4, 1/4 & 380 - 400 \\
5 & 22 - 30 & 3/10, 2/5, 3/10 & 380 - 400 \\
\hline
\end{tabular}
\end{table}

\noindent \textbf{Setting 4:} All the settings tested so far consists of networks that have a clear separation between communities, in a sense that the intra link probabilities are significantly larger than the inter link probabilities. However, there maybe scenarios where there is no clear separation between communities. In other words, the intra link probabilities have values that are close to the inter link probabilities. This setting mimics the setup of Setting 1, but with a more sparse setting and the separations between communities are less clear. The specifications of this setting are listed in Table~\ref{Setting4}.
\begin{table}[htbp]
\tiny
\centering
\caption{Specifications for Setting 4.}
\label{Setting4}
\begin{tabular}{c|c|c|c|c}
\hline
Segment & $t$ & Community Size Ratio & Link Probability & \# of Nodes\\
\hline
1 & 1 - 5 & 1/3, 1/3, 1/3 & $P(W) = 0.7, P(B) = 0.6$ & 380 - 400 \\
2 & 6 - 9 & 3/4, 1/4 & $P(W) = 0.2, P(B) = 0.1$ & 380 - 400 \\
3 & 10 - 16 & 1/4, 1/4, 1/4, 1/4 & $P(W) = 0.5, P(B) = 0.3$ & 380 - 400 \\
4 & 17 - 22 & 1/2, 1/2 & $P(W) = 0.2, P(B) = 0.1$ & 380 - 400 \\
5 & 23 - 25 & 1/5, 1/5, 1/5, 1/5, 1/5 & $P(W) = 0.4, P(B) = 0.15$ & 380 - 400\\
6 & 26 - 30 & 1/2, 1/2 & $P(W) = 0.7, P(B) = 0.55$ & 380 - 400\\
\hline
\end{tabular}
\end{table}

\noindent \textbf{Setting 5:} This setting is similar to Setting 4. But instead of having fixed link probabilities for all networks within the same segment, the probabilities are allowed to vary according to some distribution. All networks consist of nodes count ranging from 380 - 400. The rest of the specifications are listed in Table~\ref{Setting5}.
\begin{table}[htbp]
\tiny
\centering
\caption{Specifications for Setting 5.}
\label{Setting5}
\begin{tabular}{c|c|c|c}
\hline
Segment & $t$ & Community Size Ratio & Link Probability\\
\hline
1 & 1 - 6 & 1/4, 1/4, 1/4, 1/4 & $P(W) \sim U(0.2, 0.3), P(B) \sim U(0.05, 0.1) $ \\
2 & 7 - 12 & 1/2, 1/2 & $P(W) \sim U(0.45, 0.55), P(B) \sim U(0.25, 0.35) $  \\
3 & 13 - 18 & 1/2, 1/4, 1/4 & $P(W) \sim U(0.15, 0.25), P(B) \sim U(0.05, 0.10) $ \\
4 & 19 - 24 & 1/3, 2/3 & $P(W) \sim U(0.4, 0.5), P(B) \sim U(0.2, 0.3) $  \\
5 & 24 - 30 & 1/4, 1/4, 1/4, 1/4 & $P(W) \sim U(0.15, 0.25), P(B) \sim U(0.05, 0.10) $ \\
\hline
\end{tabular}
\end{table}

\noindent \textbf{Settings 6:} To test the robustness of the proposed method under misspecification, Setting 6 involved networks with correlated edges. Such networks have been studied by \cite{Saldana15}. In their paper, the parameter $\rho$ controls the correlation between network edges. The correlation used here was $\rho = 0.7$, with a dense setting. The specifications of this setting are listed in Table~\ref{Setting6}.

\begin{table}[htbp]
\centering
\caption{Specifications for Setting 6.}
\label{Setting6}
\begin{tabular}{c|c|c|c}
\hline
Segment & $t$ & Community Size Ratio & \# of Nodes\\
\hline
1 & 1 - 5 & 1/2, 1/2 & 380 - 400 \\
2 & 6 - 11 & 1/3, 1/3, 1/3 & 380 - 400 \\
3 & 12 - 19 & 3/4, 1/4 &  380 - 400 \\
4 & 20 - 24 & 1/2, 1/2 & 380 - 400 \\
5 & 24 - 25 & 3/4, 1/4 & 380 - 400 \\
6 & 26 - 30 & 2/5, 1/5, 2/5 & 380 - 400 \\
\hline
\end{tabular}
\end{table}



\ifCLASSOPTIONcaptionsoff
  \newpage
\fi



%

\bibliographystyle{IEEEtran}
\bibliography{tleeref}


%


\begin{IEEEbiographynophoto}{Rex C. Y. Cheung}
\end{IEEEbiographynophoto}

\begin{IEEEbiographynophoto}{Alexander Aue}
\end{IEEEbiographynophoto}

\begin{IEEEbiographynophoto}{Seungyong Hwang}
\end{IEEEbiographynophoto}

\begin{IEEEbiographynophoto}{Thomas C. M. Lee}
\end{IEEEbiographynophoto}




\end{document}